
\input phyzzx

\voffset=24pt
\overfullrule=0pt
\scrollmode

\hfill\vbox{\hbox{TIFR/TH/92-16}\hbox{March, 1992}}

\title{THE SPECTRUM OF $SL(2, R)/U(1)$ BLACK HOLE CONFORMAL FIELD THEORY}

\author{Dileep P. Jatkar\foot{e-mail address: dileep@tifrvax.bitnet}}

\address{Tata Institute of Fundamental Research, Homi Bhabha Road, Bombay
400005, India}

\abstract

We study string theory in the background of a two-dimensional black hole
which is described by an $SL(2, R)/U(1)$ coset conformal field theory. We
determine the spectrum of this conformal field theory using supersymmetric
quantum mechanics and give an explicit form of the vertex operators in terms of
the Jacobi functions. We also discuss the applicability of SUSY quantum
mechanics techniques to non-linear $\sigma$-models.

\endpage

\def\tL{\theta_{L}}
\def\tR{\theta_{R}}
\def\o{\omega}
\def\A{{\cal A}}
\def\ot{{1\over 2}}

\NPrefs
\def\define#1#2\par{\def#1{\Ref#1{#2}\edef#1{\noexpand\refmark{#1}}}}
\def\con#1#2\noc{\let\?=\Ref\let\<=\refmark\let\Ref=\REFS
         \let\refmark=\undefined#1\let\Ref=\REFSCON#2
         \let\Ref=\?\let\refmark=\<\refsend}

\define\BREKZ
E. Brezin and V. A. Kazakov, Phys. Lett. {\bf B236}(1990), 144.

\define\DOUSH
M. R. Douglas and S. H. Shenker, Nucl. Phys. {\bf B335}(1990), 635.

\define\GROMI
D. Gross and A. A. Migdal, Phys. Rev. Lett. {\bf 64}(1990), 127.

\define\BKZ
E. Brezin, V. A. Kazakov and Al. B. Zamolodchikov, Nucl. Phys. {\bf
B338}(1990), 673.

\define\GMIL
D. Gross and N. Milkovic, Phys. Lett. {\bf B238}(1990), 217.

\define\GIZJ
P. Ginsparg and J. Zinn-Justin, Phys. Lett. {\bf B240}(1990), 333.

\define\POL
J. Polchinski, Nucl. Phys. {\bf B346}(1990), 253.

\define\DASJEV
S. R. Das and A. Jevicki, Mod. Phys. Lett. {\bf A5}(1990), 1639.

\define\MSW
G. Mandal, A. M. Sengupta and S. Wadia, Mod. Phys. Lett. {\bf A6}(1991), 1685.

\define\WIT
E. Witten, Phys. Rev. {\bf D44}(1991), 314.

\define\GIV
A. Giveon, Mod. Phys. Lett. {\bf A6}(1991), 2843.

\define \DVV
R. Dijkgraaf, E. Verlinde and H. Verlinde, Nucl. Phys.{\bf B371}(1992), 269.

\define\DINEL
J. Distler and P. Nelson, Preprint PUPT-1262, UPR-0462T.

\define\DLP
L. Dixon, J. Lykken and M. Peskin, Nucl. Phys. {\bf B325}(1989), 329.

\define\BN
I. Bars and D. Nemeschansky, Nucl. Phys. {\bf B348}(1991), 89.

\define\B
I. Bars, Nucl. Phys. {\bf B334}(1990), 725.

\define\WITT
E. Witten, Nucl. Phys. {\bf B188}(1981), 513.

\define\GEN
L. Gendenstein, JETP Lett. {\bf 38}(1983), 356.

\define\DKS
J. W. Dabrowska, A. Khare and U. P. Sukhatme, J. Phys. {\bf A21}(1988), L195.

\define\ROC
M. Rocek, K. Schoutens and A. Sevrin, Phys. Lett. {\bf B265}(1991), 303.

\define\EFR
S. Elitzur, A Forge and E. Rabinovici, Nucl. Phys. {\bf B359}(1991), 581.

\define\BCR
K. Bardacki, M. Crescimannu and E. Rabinovici, Nucl. Phys. {\bf
B344}(1990), 344.

\define\KIR
E. B. Kiritsis, Mod. Phys. Lett. {\bf A6}(1991), 2871.

\define\AS
A. Sen, Phys. Lett. {\bf B267}(1991), 33.

\define\ASEN
A. Sen, Preprint TIFR/TH/91-37.

\define\AF
S. F. Hassan and A. Sen, Preprint TIFR/TH/91-40.

\define\KK
S. P. Khastgir and A. Kumar, Mod. Phys. Lett. {\bf A6}(1991), 3365.

\define\VEN
G. Veneziano, Phys. Lett. {\bf B265}(1991), 287.

\define\MV
K. A. Meissner and G. Veneziano, Phys. Lett. {\bf B267}(1991), 33.

\define\HH
J. H. Horne and G. T. Horowitz, Preprint UCSBTH-91-39.

\define\GR
A. Giveon and M. Rocek, Preprint IASSNS-HEP-91/84, ITP-SB-91-67.

\define\KW
V. Kac and M. Wakimoto, Proc. Nat. Acad. Sci. {\bf 85}(1988), 4956.

\define\MP
S. Mukhi and S. Panda, Nucl. Phys. {\bf B338}(1990), 263.

\define\CGK
F. Cooper, J. Ginnochio and A. Khare, Phys. Rev. {\bf D36}(1988), 2458.

\define\CFMP
C. G. Callan, D. Friedan, E. Martinec and M. Perry, Nucl Phys. {\bf
B262}(1985), 593.

\define\SMS
A. Sen, Phys. Rev. {\bf D32}(1985), 2162.

\define\SMAS
A. Sen, Phys. Rev. Lett. {\bf 55}(1985), 1846.

\chapter{Introduction}

Matrix models remain by far the most successful approach for studying string
theories in non-critical space-time dimensions\con\BREKZ\DOUSH\GROMI\noc.
We expect to gain some insight into `realistic' string theories by studying
these toy models. Of all the models, $c=1$ conformal field theory (CFT) coupled
to two dimensional gravity is the most interesting and
intriguing\con\BKZ\GMIL\GIZJ\noc.  This model can be considered as a
string theory with a two dimensional target space.\con\POL\DASJEV\noc In
the language of matrix models, these two dimensions correspond to the time
and eigenvalue of the matrix. From the continuum field theory
point of view they correspond to the $c=1$ scalar field and the Liouville
field. Using the $\sigma$-model representation of this two dimensional
theory, it was shown by Mandal et al.\MSW\ that the solution of
$\beta$-function equations in the graviton-dilaton sector described the
space-time exterior to the horizon of the black hole. Witten observed
that this solution possesses all the features of the black hole geometry.
He also showed that it is possible to construct an exact conformal
field theory\WIT\ based on an $SL(2,R)/U(1)$ gauged WZW model whose target
space has black hole geometry.(For related work see refs.\con\ROC\EFR\BCR
\KIR\noc) It was possible to get black holes with both
Euclidean and Minkowski signatures. In the gauged WZW model, this
signature depends upon which subgroup is gauged. There has been a lot of
activity since then and both Euclidean and Minkowski black hole
conformal field theories have been studied in great
detail\con\GIV\DVV\VEN\MV\AS\ASEN\AF\KK\HH\DINEL\GR\noc.

In this paper, we will study the spectrum of an $SL(2,R)/U(1)$ coset model.
We will reduce the problem of finding the spectrum to a quantum mechanical
problem in the target space. We shall determine the spectrum of this
conformal field theory for a generic value of the current algebra level
$k$, - i.e., at a generic value of central charge $c$. Even in the case of
$c=26$ conformal field theory, we shall study the full spectrum of the
conformal field theory without restricting ourselves to $(1,1)$ operators.
In the language of string theory, this means that we will study the
off-shell states as well. In general, coset models of non-compact
symmetry groups are non-unitary. But there were some suggestions that
unitary CFT's can be obtained if the spectrum is truncated by restricting
the values of $l$ to $-1/2 > l > -k/2$\con\DLP\BN\B\noc.

For orientation and motivation, we shall review the $SL(2,R)/U(1)$ coset
model in section 2, and show  that the problem of
determining the spectrum of Virasoro primary fields can be reduced to a
quantum mechanical problem in the target space. We
shall then introduce the techniques of supersymmetric (SUSY) quantum mechanics
in section 3, and
briefly discuss the concept of shape invariance and its relation with SUSY
quantum mechanics. In section 4, we shall apply these techniques to the problem
at hand. We shall show, using shape invariance and SUSY, that the black hole
quantum mechanics problem is exactly soluble. We give the exact bound state as
well as scattering spectrum and write down the wave-functions explicitly.
Utilising the relation of this quantum mechanical problem with black hole
CFT, we show that the bound state spectrum gives the conformal dimensions
of the vertex operators which are the eigenfunctions themselves. Some of
these results were obtained by Dijkgraaf, Verlinde and Verlinde\DVV\ using
different techniques. In section
5, we shall study the possibility of wider applicability of this technique.
In particular, we shall show that this method can be applied to CFT with
$\sigma$-model representation.

\chapter{Review of $SL(2,R)/U(1)$ coset model}

The black hole CFT, as shown by Witten\WIT, can be formulated as a gauged
WZW model. The WZW model is based on a non-compact symmetry group
$SL(2,R)$.  The symmetry that is gauged in this model corresponds to some
abelian subgroup $H$ of $SL(2,R)$. When $H$ is compact, we get a Euclidean
black hole target space whereas for a non-compact $H$, we get a Minkowski
black hole. Since we shall concentrate on Euclidean black hole quantum
mechanics, we shall only review the Euclidean version of the black hole CFT
\foot{In this section we shall follow the analysis of ref.\DVV .}.

Let us parametrize the $SL(2,R)$ group manifold by three real coordinates
$r$, $\tL$ and $\tR$. $\tL$
and $\tR$ are periodic coordinates -i.e. they are compact-, whereas $r$ is
non-compact and takes values on the non-negative real axis. We write the
field on the group manifold as
$$
g = \exp{({i\over 2}\tL\sigma_{2})}\exp{({1\over
2}r\sigma_{1})}\exp{({i\over 2}\tR\sigma_{2})}\eqn\rone
$$
where $\sigma_{i}$ are the Pauli matrices. The abelian subgroup $H$ is
generated by $\sigma_{2}$, and the gauge transformation is a shift symmetry in
$\tL$ and $\tR$($\theta_{L,R}\rightarrow \theta_{L,R} + \alpha$). With this
parametrization of $g$, the gauged WZW action is given by
$$
\eqalign{S &= S_{WZW}[r,\tL\ ,\tR\ ] + {k\over 2\pi}\int d^{2}z
[A(z,\bar z)
(\bar\partial\tR + \cosh{r}\bar\partial\tL)\cr &+ \bar A(z,\bar z)
(\bar\partial\tL + \cosh{r}\bar\partial\tR) - A(z,\bar z)\bar A(z,\bar z)
(\cosh{r} + 1)]}\eqn\rtwo
$$
where
$$
S_{WZW}[r,\tL\ ,\tR\ ] = {k\over 4\pi}\int d^{2}z(\bar\partial r\partial
r - \bar\partial\tL\partial\tL - \bar\partial\tR\partial\tR - 2\cosh{r}
\bar\partial\tL\partial\tR).\eqn\rthree
$$
Gauge fixing can be done by parametrizing the gauge field as
$$
\eqalign{A &= \partial\phi_{L}\cr \bar A &= \bar\partial\phi_{R}\cr}\eqn\rthpr
$$
where $\phi_{L}$ and $\phi_{R}$ are complex fields with the condition
$\phi_{L} = (\phi_{R})^{*}$. We are assuming a trivial world sheet topology
while writing eq.\rthpr. If we shift $\tL\rightarrow \tL + \phi_{L}$ and
$\tR\rightarrow \tR + \phi_{R}$ in the action \rtwo\ and use the gauge
invariance, we see that the action depends
only on the difference $\phi = \phi_{L} -\phi_{R}$.  The gauge fixed action
is given by
$$
S_{gf} = S_{WZW}[r, \tL\ ,\tR\ ] + S[\phi\ ]
+ S[b,c]\eqn\rfour
$$
where $S[\phi]$ is the action of a time-like free
scalar field and $S[b,c]$ describes the spin $(1,0)$ ghost system. The stress
energy tensor of the coset model can be written as
$$
T(z) = {1\over k -
2}\eta_{ab}J^{a}J^{b} + {k\over 4}(\partial\phi)^{2} + b\partial
c\eqn\rfive
$$
where $\eta_{ab}$ is the metric on the $SL(2,R)$ Lie algebra
and $J^{a}$ are $SL(2,R)$ currents. The scalar field $\phi$ is
compactified in the case of a Euclidean black hole.
Since the gauge fixed theory
contains a free scalar field theory and an ungauged $SL(2,R)$ WZW model, the
vertex operators of the coset model are products of the vertex operators of
the
free scalar field theory and the $SL(2,R)$ WZW model -i.e.,
$$
V(z,\bar z) = T(r(z,\bar z),\tL(z,\bar z), \tR(z,\bar
z))\exp{(iq_{L}\varphi + iq_{R}\bar\varphi)}\eqn\rsix
$$
where $\phi = \varphi + \bar\varphi$.
Vertex operators of the gauged WZW model should satisfy the constraint
$J^{3} -\bar J^{3} = 0$. In the formulation given above, this constraint is
imposed by the BRST charge
$$
Q_{B} = \int dz c(J^{3}+{i\over 2}k\partial\phi) + c.c\eqn\rsp
$$
The zero modes of the $SL(2,R)$ currents act as differential operators on
the vertex operators
$$
\eqalign{J^{3} &= -i{\partial\over \partial\tL}\cr J^{\pm} &= \exp{(\pm\
i\tL)} ({\partial\over\partial r} \mp {i\over \sinh{r}}
({\partial\over\partial\tR} - \cosh{r}{\partial\over\partial\tL})).}\eqn\rseven
$$
Using the Sugawara relation, the Virasoro generators can be written in terms of
the modes of $SL(2,R)$ currents and the abelian current. We wish to find
the full spectrum of the Virasoro primary fields. But we shall first
concentrate on the primary fields of the current algebra. They correspond to
vertex operators without any oscillator excitations. To determine the spectrum
of these primary fields using the Virasoro generator $L_{0}$, it suffices to
concentrate only on the zero mode bilinears of currents in the Sugawara
relation
$$
L_{0} = {1\over k-2}\eta_{ab}J^{a}_{0}J^{b}_{0}-{1\over
k}p_{0}^{2}\eqn\rnseven
$$
where $p_{0}$ is the momentum conjugate to the zero mode $\phi_{0}$. The
Virasor
   o
generator $L_{0}$ expressed in terms of zero modes of the coordinates is
$$
L_{0} = -{\Delta_{0}\over k - 2} - {1\over k}{\partial^{2}\over\partial\tL^{2}}
\eqn\reight
$$
where
$$
\Delta_{0} = {\partial^{2}\over\partial r^{2}} +
\coth{r}{\partial\over\partial r} + {1\over \sinh^{2}{r}}
({\partial^{2}\over\partial\tL^{2}} -
2\cosh{r}{\partial^{2}\over\partial\tL\partial\tR} +
{\partial^{2}\over\partial\tR^{2}})\eqn\rnine
$$
is the $SL(2,R)$ Casimir operator. Thus the equation
$$
L_{0}V = hV\eqn\rten
$$
becomes a Schr\" odinger-like equation with $V$ being
the eigenfunctions of the operator $L_{0}$ with eigenvalues $h$. As was
mentioned earlier, we shall
consider off-shell string modes but with the constraint $L_{0} - \bar  L_{0}
= 0$. This constraint enables us to decompose $T(r,\tL\ ,\tR)$
into $T(r,\theta)$ and $T(r,\tilde\theta)$ (where $\theta = (\tL + \tR)/2$ and
$\tilde\theta = (\tL -\tR)/2$) which are the momentum and
winding tachyons. In terms of the new variables, the $L_{0}$ operator for the
tachyon field $T(r,\theta)$ is
$$
L_{0} = - {1\over k - 2}\left[{\partial^{2}\over\partial r^{2}} +
\coth{r}{\partial\over\partial r} + (\coth^{2}({r\over 2}) - {2\over
k}){\partial^{2}\over\partial\theta^{2}}\right].\eqn\relvn
$$
On the other hand, $L_{0}$ for $T(r,\tilde\theta)$ becomes
$$
L_{0} = - {1\over k - 2}\left[{\partial^{2}\over\partial r^{2}} +
\coth{r}{\partial\over\partial r} + (\tanh^{2}({r\over 2}) - {2\over
k}){\partial^{2}\over\partial\tilde\theta^{2}}\right].\eqn\rtwel
$$
The group invariant measure inherited by the coset model from $SL(2,R)$
group manifold is
$$
dg = {1\over 4\pi^{2}}\sinh{r}drd\theta.\eqn\rthrtn
$$
Therefore, the inner product of the tachyons $T_{1}$ and $T_{2}$ on the target
space is defined as
$$
\langle T_{1}|T_{2}\rangle = \int dg T_{1}(r,\theta)T_{2}(r,\theta).\eqn\rfortn
$$
We linearise the integration measure by absorbing $\sinh^{1/2}(r)$ in
$T(r,\theta)$. This redefinition further simplifies the form of the $L_{0}$
operator and reduces it to the Schr\" odinger operator given by
$$
{1\over k - 2}(-{\partial^{2}\over\partial r^{2}} + V(r)),\eqn\rfiftn
$$
where $V(r)$ for the winding states is given by
$$
V(r) = (\o^{2} - {1\over 16})\tanh^{2}({r\over 2}) - {1\over 16}
\coth^{2}({r\over 2}) - {2\o^{2}\over k} + {3\over 8}\eqn\rsixtn
$$
and for the momentum states is given by
$$
V(r) = (\o^{2} - {1\over 16})\coth^{2}({r\over 2}) - {1\over 16}
\tanh^{2}({r\over 2}) - {2\o^{2}\over k} + {3\over 8}.\eqn\rsevntn
$$
We have replaced $\partial^{2}/\partial\theta^{2}$
($\partial^{2}/\partial\tilde\theta^{2}$) by $-\o^{2}$. We shall
carry out our analysis for an arbitrary value of $k$. The spectrum
of the black hole problem can be obtained by putting $k=9/4$.
Thus we see that solving the problem $L_{0}T = hT$ is reduced to solving a
quantum mechanical problem with a specific potential. This way we can
determine only those Virasoro primary fields which are also the current
algebra primary fields. The
remaining Virasoro primaries can be determined from the spectrum of these
primary fields by attaching a string of currents and its derivatives on
the left. The condition that these new vertex operators are physical
fields of the coset model gives constraints on the string of currents.
As a result of this constraint, only those combinations of the currents and
its derivatives which commute with the BRST charge are allowed. In the next
section we will discuss the techniques of SUSY quantum mechanics and shape
invariance which will be used to show that the problem
stated above is exactly solvable.

\chapter{SUSY Quantum Mechanics and Shape Invariance}

The concept of supersymmetry in quantum mechanics was introduced by Witten
\WITT . Supersymmetry, as is well known, relates bosons to fermions. In
quantum mechanics two Hamiltonians which are related to each other by
supersymmetry are called partner Hamiltonians. These Hamiltonians, due to
SUSY, are isospectral except for the ground state. In short, SUSY in
quantum mechanics can be described as follows. Given a potential
$V_{-}(x)$, SUSY allows us to construct a partner potential $V_{+}(x)$
which has an identical energy spectrum except for the ground state. Such a
pair of Hamiltonians is given by
$$
H_{\pm} = - {d^{2}\over dx^{2}} + V_{\pm}(x)\eqn\sone
$$
Using two component notation, a Hamiltonian $\cal H$ can be written as an
anticommutator of SUSY charges as follows
$$
{\cal H} = \pmatrix{H_{-}&0\cr 0&H_{+}\cr} = \ot\{ Q,Q\}.\eqn\stwo
$$
The supersymmetry charge $Q$ commutes with the Hamiltonian $\cal H$. If we
parametrize the SUSY charge $Q$ as
$$
Q = \pmatrix{0&\A^{\dag}\cr \A&0\cr}\eqn\sthree
$$
then both $H_{+}$ and $H_{-}$ can be written in the following factorised forms
$$
H_{+} = \A\A^{\dag}\qquad {\rm and}\qquad H_{-} = \A^{\dag}\A.\eqn\sfour
$$
For the choice
$$
\A = {d\over dx} + W(x)\eqn\sfive
$$
the partner potentials can be written in terms of the superpotential
$W(x)$
$$
V_{\pm}(x) = W^{2}(x) \pm {dW(x)\over dx}\eqn\ssix
$$
Now, it trivially follows that when SUSY is unbroken, the ground state of
$H_{-}$ has zero energy and the ground state eigenfunction is
$$
\Phi^{-}_{0}(x) = N_{0}\exp{(-\int^{x} W(x')dx')}.\eqn\sseven
$$
The partner Hamiltonians have an identical bound state spectrum except for
the ground state of $H_{-}$ so that
$$E^{-}_{n+1} = E^{+}_{n} \qquad n = 0,1,2,\cdots .\eqn\snsevn
$$
The eigenfunctions corresponding to
any given eigenvalue are  related to each other through $\A$ and
$\A^{\dag}$ as follows
$$
\eqalign{\A\Phi^{-}_{n+1}(x) &= (E^{+}_{n})^{1/2}\Phi^{+}_{n}(x)\cr
\A^{\dag}\Phi^{+}_{n}(x) &= (E_{n}^{+})^{1/2}\Phi^{-}_{n+1}(x).}\eqn\seight
$$
Since SUSY does not give the eigenvalues and eigenfunctions corresponding
to excited states we need some additional information to actually
determine the spectrum. To determine the eigenvalues and eigenfunctions we
need to invoke the concept of shape invariance. The term shape invariance
was introduced by Gendenstein\GEN . If the potentials $V_{+}$ and $V_{-}$
have a similar shape, they are called shape invariant potentials.
In mathematical terms, the shape invariance condition can be written as
the Ricatti equation
$$
W^{2}(x,a_{i}^{[0]}) + {dW(x,a_{i}^{[0]})\over dx} = \tilde W^{2}(x,
a_{i}^{[1]})- {d\tilde W(x,a_{i}^{[1]})\over dx} + c(a_{i}^{[1]})\eqn\snine
$$
where $a_{i}^{[n]}$ are the parameters appearing in the potential. If the
an analytical solution to this Ricatti equation exists, $i.e.$, if $\tilde
W$ can be expressed in a closed form\foot{Throughout this paper, SUSY
implies that the superpotential $W$ can be determined analytically and can
be written in a closed form. Similarly, shape invariance means that the
Ricatti equation can be solved analytically.}, then by using SUSY, we can
completely determine the spectrum of eigenvalues and eigenfunctions by
purely algebraic means. This can easily be seen as follows. On the left
hand side of the Ricatti equation we have the potential $V_{+}$. As we saw
earlier, SUSY does not give us any information about the ground state of
this potential. But, since right hand side of the Ricatti equation is
written in terms of $\tilde V_{-}$, we can find the ground state
eigenvalue and eigenfunction using SUSY. From the Ricatti equation it is
easy to see that the spectrum of $V_{+}$ and $\tilde V_{-}$ are identical
upto an overall shift in the eigenvalues. Thus the zero energy SUSY ground
state of $\tilde V_{-}$ is the ground state of $V_{+}$ with the energy
$c(a_{i}^{[1]})$. Now, recall that $V_{+}$ has one state less than
$V_{-}$ and that is the ground state of $V_{-}$. In other words, the
ground state eigenvalue of $V_{+}$ is equal to the first excited state
eigenvalue of $V_{-}$. Now if we can find the parameters $a_{i}^{[n]}$ as a
function $f(a_{i}^{[n-1]})$ of $a_{i}^{[n-1]}$ such that the Ricatti
equation \snine\ is satisfied, then following the earlier discussion, the
ground state of $V_{-}(x,a_{i}{[n]})$ has the same energy as the first
excited state of $V_{-}(x,a_{i}^{[n-1]})$. This state in turn has the same
energy as the second excited state of $V_{-}(x,a_{i}^{[n-2]})$ and so on.
Thus we see that $V_{-}(x,a_{i}^{[n]})$ gives a potential for each
parameter set $a_{i}^{[n]}$ and it is easy to see that $V_{-}(x,a_{i}^{[n]})$
has the same spectrum as that of $V_{-}(x,a_{i}^{[0]})$ except
that the lowest $n$ states of $V_{-}(x,a_{i}^{[0]})$ are missing.(For
details and specific examples see ref.\CGK.) Since
SUSY intertwines these potentials it is possible to determine the excited
state eigenfunctions for any of these problems using eq.\seight. Thus we
can determine the full spectrum of eigenvalues and eigenfunctions.
Thus using shape invariance we can show that
the energy spectrum of $H_{-}$ is given by
$$
E^{(-)}_{n} = \sum_{k=1}^{n} c(a_{i}^{[k]}).\eqn\sten
$$
The eigenfunctions can be determined by using
eq.\seight\ for successive values of parameters $a_{i}^{[n]}$, $i.e.$
$$
\Phi^{(-)}_{n}(x,a_{i}^{[0]}) = \A^{\dag}(x,a_{i}^{[0]})\A^{\dag}(x,
a_{i}^{[1]})\cdots \A^{\dag}(x,a_{i}^{[n-1]})\Phi_{0}^{(-)}(x,
a_{i}^{[n]}).\eqn\seleven
$$

\chapter{Black Hole Quantum Mechanics}

In this section we shall apply the techniques developed in the previous
section to the black hole problem. Recall that in sec. 2, we reduced the
black hole CFT problem to a quantum mechanics problem on a half line.

First, let us note that the spectrum of the potential in eq.\rsixtn\ contains
both bound states and scattering states, whereas the potential in eq.\rsevntn\
does not have any bound state for $\o^{2} > 1/16$. We shall see in our
analysis that even for $\o^{2} < 1/16$ and $\o$  real, this potential does
not have any bound state.

\section{Winding Sector}

In this subsection we shall study only the winding sector $i.e.$, the
potential given in eq.\rsixtn . We shall first show that this potential has
supersymmetry and use it to determine the superpotential. Then we shall
invoke the shape invariance property of this potential to find out the
complete spectrum of bound state eigenvalues and eigenfunctions. The
scattering spectrum is determined by analytically continuing the bound state
spectrum. From the asymptotic behaviour of these analytically continued
wave-functions, we shall derive the scattering matrix and subsequently the
density of states. The potential in eq.\rsixtn\ can be obtained from the
superpotential
$$
W(r) = (-{1\over 4}\pm\o)\tanh {({r\over 2})} - {1\over 4}\coth{({r\over
2})}.\eqn\bone
$$
Therefore, the Schr\" odinger equation written in terms of the
superpotential is given by
$$
\left( -{d^{2}\over dr^{2}} + W^{2}(r) - {dW(r)\over dr} \pm\o  -
{2\o^{2}\over k}\right)\Phi = E\Phi.\eqn\btwo
$$
Since from SUSY quantum mechanics we know that
$$
\left( -{d^{2}\over dr^{2}} + W^{2}(r) - {dW(r)\over dr}\right)\Phi_{0} =
0\eqn\bntwo
$$
where $\Phi_{0}$ is the ground state eigenfunction, the ground state energy
in eq.\btwo\ is
$$
E_{0} = \pm\o - {2\o^{2}\over k}.\eqn\bthree
$$
The ground state wave-function is given by
$$
\eqalign{\Phi_{0}(r) &= \exp{(-\int^{r} W(r')dr')} =
\exp{(-\int^{r}({1\over 4}\pm\o)\tanh{({r\over 2})}dr + \int^{r}{1\over
4}\coth({r\over 2})dr)}\cr &= {\sinh^{1/2}(r)\over \sqrt{2}(\cosh({r\over
2}))^{\pm 2\o}}.}\eqn\bfour
$$
The condition for this wave-function\foot{Recall that $\sinh^{1/2}(r)$ is
precisely the prefactor that we absorbed in the wave-function to linearise
the integration measure. So the wave-function corresponding to the
original problem is without the $\sinh^{1/2}(r)$ term. In addition, it
will have a $\theta$ dependent part as well. But this will not
affect any of our conclusions.} to be a bound state wave-function ($i.e.$
square integrable) is $\pm\o > 1/2$. Since $\o$ takes both positive and
negative values we can write this condition as $|\o| >1/2$. We
shall come back to this point later in this section when we will discuss
square integrability of the full spectrum of bound state eigenfunctions.

Now we shall invoke the shape invariance property to determine the complete
spectrum of eigenvalues and eigenfunctions. Let us show that the
potential $V(r)$ given in eq.\rsixtn\ is shape invariant. In fact, this
can be shown for any general potential of the type given in eq. \rsixtn,
$i.e.$, we shall consider a potential with the same functional form but
with arbitrary parameters in front of the $\tanh^{2}(r/2)$ and the
$\coth^{2}(r/2)$ instead of $(\o^{2}-1/16)$ and $(-1/16)$. Also instead of
$r/2$ we shall use $\alpha r$. Let us choose\DKS
$$
W(r,a_{i}^{[0]}) = A\tanh(\alpha r) - B\coth(\alpha r)\eqn\bfive
$$
with the parameter set $a_{i}^{[0]} = (A,B)$. Then it follows that
$$
\eqalign{V_{-}(r,a_{i}^{[0]}) &= W^{2}(r,a_{i}^{[0]}) - {dW(r,
a_{i}^{[0]})\over dr}\cr &= (A^{2} + A\alpha)\tanh^{2}(\alpha r) + (B^{2} -
B\alpha)\coth^{2}(\alpha r) -2AB + (A - B)\alpha}\eqn\bsix
$$
and
$$
\eqalign{V_{+}(r,a_{i}^{[0]}) &= W^{2}(r,a_{i}^{[0]})+{dW(r,a_{i}^{[0]})\over
dr}\cr &= (A^{2} - A\alpha)\tanh^{2}(\alpha r) + (B^{2} +
B\alpha)\coth^{2}(\alpha r) -2AB - (A - B)\alpha.}\eqn\bnsix
$$
As we have seen earlier in sec. 3, to determine the first excited state
eigenvalue and eigenfunction of $V_{-}$, we need to know the ground state
eigenvalue and eigenfunction of $V_{+}$ which
can be determined using the Ricatti equation. Now if we choose $a_{i}^{[1]}
= (A-\alpha\ , B+\alpha)$, the new superpotential is given by
$$
\tilde W(r,a_{i}^{[1]}) = (A-\alpha)\tanh(\alpha r) - (B+\alpha)coth(\alpha
r).\eqn\bseven
$$
With this choice of superpotential it is easy to see that eq.\snine\ is
satisfied and the constant $c(\{ a_{1}\}) = 4\alpha(A-B-\alpha)$. Thus
we see that energy of the first excited state is $4\alpha(A-B-\alpha)$. We
can also determine the ground state wave-function of the partner potential
$V_{+}(r)$ and as a consequence of eq.\seleven, the first excited state
wave-function of $V_{-}(r)$. In fact the Ricatti equation \snine\ in this
case can be solved by doing successive transformations and choosing
appropriate $a_{i}^{[n]} = f(a_{i}^{[n-1]})$. The form of the function remains
the same throughout. The set $a_{i}^{[n]}$ in the above example is
$(A-n\alpha , B+n\alpha)$, and therefore the energy spectrum is
$$
E_{n} = 4n\alpha(A-B) - 4n^{2}\alpha^{2}.\eqn\beight
$$
The eigenfunctions can be deduced by successively using eq.\seleven\ and are
given by
$$
\Phi_{n}(y) = (y-1)^{B/2\alpha}(y+1)^{-A/2\alpha}
P_{n}^{({B\over\alpha}-\ot, -{A\over\alpha}-\ot)}(y)\eqn\bnine
$$
where $y = \cosh 2\alpha r$ and $P_{n}^{(\beta\ ,\gamma)}(y)$ are Jacobi
functions.

Comparing eq.\bsix\ and \rsixtn, we see that for the winding mode sector,
$A = (|\o|-1/4)$, $B = 1/4$ and $\alpha = 1/2$. Since in our problem the
ground state energy is not zero, the whole spectrum gets shifted. Substituting
the values of $A$, $B$ and $\alpha$ and after adding the ground state
energy in eq.\bthree\ to the spectrum \beight, the energy eigenvalues are
$$
E_{n} = (2n+1)|\o| - n(n+1) - {2\o^{2}\over k}\eqn\bten
$$
and the eigenfunctions are
$$
\Phi_{n}(y) = (y-1)^{1/4}(y+1)^{-|\o|+1/4}P_{n}^{(0,-2|\o|)}(y)\eqn\beleven
$$
where $y=\cosh(r)$. The square integrability of these eigenfunctions,
by virtue of eq.\seleven, is determined by the square integrability of the
ground state eigenfunction of the $n$th problem($i.e.$ the ground state
eigenfunction of $V_{-}(r,\{ a_{n}\})$). For the potential with
parameters $a_{i}^{[n]} = (A-n\alpha\ , B+n\alpha)$, the condition for
square integrability of the ground state wave-function for any $n$, as can be
deduced from eq.\bfour, depends on the difference of these two parameters,
$i.e.$ it reduces to $A-B >0$.  Substituting the values of $A$ and $B$ we
get the relation $|\o|-n > 1/2$. Thus we see that for a fixed value of
$\o$, there always exist a finite number of bound states.

Now let us turn our attention to the $SL(2,R)$ Casimir operator. In the
winding sector, the Casimir operator $\Delta$ takes the form
$$
\Delta = {\partial^{2}\over\partial r^{2}} + \coth r{\partial\over\partial
r} + (\tanh^{2}{r\over 2} -
1){\partial^{2}\over\partial\tilde\theta^{2}}.\eqn\btwelve
$$
Comparing the spectra of the Casimir operator $\Delta$ and that of the
Virasoro generator $L_{0}$, we see that
$$
\eqalign{{\rm Spec}(\Delta) &= {\rm Spec}[(k-2)(-L_{0}-{2\o^{2}\over
k}+\o^{2})]\cr &= (-2n+1)|\o| + n(n+1) + \o^{2}.}\eqn\bthrtn
$$
Equating the right hand side of eq.\bthrtn\ with the usual eigenvalue of
$l(l+1)$ the Casimir, we get the relation $l+|\o| = n$. Thus, from
this relation we see that the number of nodes of the bound state
wave-function are related to the sum $l+|\o|$. Using this relation along
with the square integrability condition, we find that bound state spectrum
corresponds to $l<-1/2$ representations. But $l<-1/2$ and $l+|\o|$ being a
non-negative integer are the properties of the discrete representations of
$SL(2,R)$. Hence, the bound state spectrum forms a discrete representation of
$SL(2,R)$.

The scattering matrix can be determined by analytically continuing the
bound state spectrum. We, therefore, express the wave-function in
terms of the hypergeometric function
$$
\Phi_{n}(y) = (y-1)^{1/4}(y+1)^{-\o
+1/4}{1\over\Gamma(n)}F(-n,1-n-|\o|,1;{1-y\over 2}).\eqn\bfortn
$$
We then analytically continue $n$ to $|\o|+i\lambda-1/2$. Substituting this
value of $n$ and taking large $y$ asymptotics of the hypergeometric
function, the asymptotic form of the wave-function can be written as
$$
\eqalign{\Phi &\sim (\ot)^{-\ot+|\o|+i\lambda}
{\Gamma(2i\lambda)\over \Gamma(\ot+|\o|+i\lambda)\Gamma({1\over
2}-|\o|+i\lambda) \Gamma(-\ot+|\o|+i\lambda)}\exp(i\lambda r)\cr &+
(\ot)^{-\ot+|\o|-i\lambda}{\Gamma(-2i\lambda)\over
\Gamma(\ot+|\o|-i\lambda)\Gamma(\ot-|\o|-i\lambda)
\Gamma(-\ot+|\o|+i\lambda)}\exp(-i\lambda r).}\eqn\bfiftn
$$
It is now easy to extract from this expression, the scattering amplitude
$$
S = 2^{-2i\lambda}{\Gamma(2i\lambda) \Gamma(\ot+|\o|-i\lambda)
\Gamma(\ot-|\o|-i\lambda)\over \Gamma(-2i\lambda) \Gamma({1\over
2}+|\o|+i\lambda) \Gamma(\ot-|\o|+i\lambda)}.\eqn\bsixtn
$$
The density of states, therefore, is given by
$$
\rho(\lambda) = 2 Re[2\psi(2i\lambda) - \psi(\ot+|\o|+i\lambda) -
\psi(\ot-|\o|+i\lambda)]\eqn\bsevntn
$$
where, $\psi(z) = \Gamma'(z)/\Gamma(z)$.

\section{Momentum Sector}

Let us consider the potential given in eq.\rsevntn. As we argued earlier
this potential blows up at $r=0$ for $|\o|>1/4$. On the other hand, when
$|\o|<1/4$ the potential goes to $-\infty$ at $r=0$ and can in principle
have bound states. But as it turns out, the ground state wave-function
$$
\Phi_{0} = \cosh^{-1/2}{r\over 2}\sinh^{\pm 2\o-1/2}{r\over 2}\eqn\beightn
$$
is not square integrable and hence, this potential does not have any bound
states for $|\o|<1/4$ either.

The scattering wave-function can be written in terms of a hypergeometric
function as
$$
\tilde\Phi = (y-1)^{1/4 \pm\o}(y+1)^{1/4}{\Gamma(\ot\pm\o
+i\lambda)\over \Gamma(-\ot+i\lambda)\Gamma(\ot+i\lambda)}
F(\ot-i\lambda, \ot\pm 2\o+i\lambda; {1-y\over 2}).\eqn\bnintn
$$
We look at the asymptotic behaviour of this wave-function to determine the
scattering amplitude. The asymptotic form of the wave-function is given by
$$
\eqalign{\tilde\Phi(r\rightarrow\infty) &\sim
(\ot)^{-\ot\pm\o+i\lambda}{\Gamma(2i\lambda)\Gamma(1\pm 2\o)\over
\Gamma(1\ot\mp\o+i\lambda)\Gamma(\ot\mp\o+i\lambda)\Gamma(\ot\pm\o+i\lambda)}
\exp(i\lambda r)\cr &\!\!\!\!\!\!\!\!\!\!\!\!\!\!\!\!\!\!+
(\ot)^{-\ot\mp\o-i\lambda}{\Gamma(-2i\lambda)\Gamma(1\pm
2\o)\Gamma(\ot\pm\o+i\lambda)\over
\Gamma(\ot\pm\o-i\lambda)\Gamma(-\ot\mp\o+i\lambda)
\Gamma(\ot\mp\o+i\lambda)\Gamma(\ot\pm\o-i\lambda)}\exp(-i\lambda r)}\eqn\btnt
$$
The ratio of the coefficients of the incoming and outgoing waves gives the
scattering matrix
$$
S = 2^{-2i\lambda}{\Gamma(2i\lambda)\Gamma(\ot\pm\o+i\lambda)
\Gamma(\ot\pm\o-i\lambda)\over\Gamma(\ot\pm\o+i\lambda)
\Gamma(-2i\lambda)\Gamma(\ot\pm\o+i\lambda)}.\eqn\btone
$$
The density of states is given by
$$
\rho(\lambda) = 4{\rm Re}[\psi(2i\lambda) -\psi(\ot\pm\o+i\lambda)]\eqn\bttwo
$$
where $\psi(z) = \Gamma '(z)/\Gamma(z)$.

In this section, we studied the spectrum of the chiral algebra primary
fields. This covers all the chiral algebra primary fields but as far as
primary conformal fields, $i.e.$, Virasoro primary fields are concerned,
this set is far from complete. The Virasoro primary fields which are not
the chiral algebra primary fields can be obtained by attaching a string of
chiral algebra currents and its derivatives to the left of the chiral
algebra primary field. The states corresponding  to these new fields will
be in the cohomology of $Q_{B}$ provided the string of chiral currents
commutes with $Q_{B}$. Apart from this, they have to satisfy
additional constraints to be the Virasoro primary fields\DINEL. Their
conformal dimension, however, can be easily read out from their
composition. In the language of states, this means that the oscillator creation
operators contribute a specific integer in addition to the conformal
dimension of the basic chiral primary field.

These fields along with the original set give the complete set of Virasoro
primary fields. Though we identify the wave-functions with the vertex
operators, it is important to recognise, at this stage, that the
wave-functions are expressed in terms of the zero modes of the fields
$r(z,\bar z)$ and $\theta(z,\bar z)$(or $\tilde\theta(z,\bar z)$). On the
other hand the vertex operators are expressed in terms of the fields $r(z,\bar
z)$ and $\theta(z,\bar z)$(or $\tilde\theta(z,\bar z)$).
Therefore to write correct expressions for the vertex operators, we
need to replace the zero modes in the wave-functions by the fields $r$ and
$\theta$ and then regularise them.

\chapter{Application to $\sigma$-models}

In the previous section, we used the techniques of SUSY quantum mechanics to
solve the black hole problem. Here we will show that this method is quite
general and is applicable to conformal field theories with $\sigma$-model
representations. To illustrate this, let us consider a non-linear
$\sigma$-model in the background of the graviton $G_{\mu\nu}$, the dilaton
$D$ and the tachyon $T$\con\SMAS\SMS\CFMP\noc. The $\sigma$-model action with
$d$ scalar fields is given by
$$ S_{\sigma} =
{1\over 4\pi\alpha'}\int d^{2}z \sqrt{g}\left({1\over 2}g^{ab}G_{\mu\nu}
\partial_{a} x^{\mu}\partial_{b} x^{\nu} - \alpha'R^{(2)}D(x) +
T(x).\right)\eqn\dnone
$$
The condition of conformal invariance of this $\sigma$-model is
implemented by setting the $\beta$-functions to zero. This condition gives
the equations of motion of $G_{\mu\nu}$, $D$ and $T$ as
$$\eqalign{
R_{\mu\nu} - 2\nabla_{\mu}\nabla_{\nu}D + \nabla_{\mu}T\nabla_{\nu}T &= 0\cr
R + 4 (\nabla D)^{2} - 4 \nabla^{2}D + (\nabla T)^{2} + V(T) + c &= 0\cr
-2\nabla^{2}T + 4\nabla D\nabla T + V'(T) &= 0}\eqn\dntwo
$$
where $c=(d-26)/3\alpha'$. These equations can be derived from the target
space action
$$
S = \int d^{d}x \exp(-2D)\sqrt{G} [ R - 4(\nabla D)^{2} + (\nabla T)^{2} +
V(T) + {26-d\over 3}].\eqn\dnthr
$$
It is well known that, in $\sigma$-model representation, the $L_{0}$
operator on the world sheet is identified with the target space Laplacian
$\Delta$. It is given by
$$
\Delta = {1\over
\exp(-2D)\sqrt{G}}\nabla_{\mu}\exp(-2D)\sqrt{G}G^{\mu\nu}\nabla_{\nu}
\eqn\done
$$
where $\nabla_{\mu}$ is the covariant derivative, $G_{\mu\nu}$ is the metric
on the target space and $D$ is the dilaton field. Thus we see that the
problem of determining the spectrum of Virasoro vertex operators is
equivalent to finding all the solutions of the Laplacian $\Delta$. For a
fixed background $G_{\mu\nu}$ and $D$, the Laplacian can be simplified.
This reduces the problem to a second order differential equation. We can
absorb the factor $G^{1/4}\exp(-D)$ in the solution, exactly in the
same way as we did in the black hole conformal field theory in sec.2. In
the case of a fixed background, it is always possible to absorb such a factor.
Since this transformation linearises the integration measure we can write the
Laplacian as a Schr\" odinger operator. It is for
this Schr\" odinger equation that the techniques of SUSY quantum
mechanics can be used, although for an arbitrary background, this problem may
not have supersymmetry.

Now let us consider an exactly solvable quantum mechanics problem given by
$$
({d^{2}\over dr^{2}} + V(r))\Psi(r) = E\Psi(r)\eqn\dtwo
$$
where $V(r)$ is a potential whose full spectrum of eigenvalues and
eigenfunctions is known. Let these wave-functions be characterised by a
set of quantum numbers $\{ b_{i}\}$. Then it is always possible to
decompose any eigenfunction $\Psi$ into a product of two functions $\phi$
and $\chi$. The function $\phi$ carries all the information about $\{
b_{i}\}$ while $\chi$ is independent of $\{ b_{i}\}$, $i.e.$
$$
({d^{2}\over dr^{2}} + V(r))\chi(r)\phi(r,\{ b_{i}\}) = E\chi(r)\phi(r,\{
b_{i}\}).\eqn\dthree
$$
If we eliminate the function $\chi(r)$ from the above equation, we get
$$
({1\over \eta(r)}{\partial\over \partial r}\eta(r){\partial\over \partial
r} + \tilde V(r))\phi(r,\{ b_{i}\}) = E \phi(r,\{ b_{i}\})\eqn\dfour
$$
where $\eta(r)$ is determined by $\chi(r)$. Let us consider, as an example,
the target space to be two dimensional. Let it be parametrized by $r$ and
$\theta$. Then we can interpret of the operator on the L.H.S. of eq.\dfour\ as
a Laplacian in this two dimensional target space. With the identification
$$
\tilde V(r) = \bar V(r){\partial^{2}\over \partial\theta^{2}}\eqn\dabc
$$
and
$$
\Phi(r) = \phi(r,\{ b_{i}\})e^{im\theta},\eqn\dfive
$$
eq.\dfour\ can be written as
$$
({1\over \eta(r)}{\partial\over \partial r}\eta(r){\partial\over \partial
r} + \bar V(r){\partial^{2}\over \partial\theta^{2}})\Phi(r,\theta) =
E\Phi(r, \theta).\eqn\dsix
$$
Now it is easy to read out the metric $G_{\mu\nu}$ and the dilaton $D$
from the Laplacian occuring in eq.\dsix. The metric is given by
$$
ds^{2} = dr^{2} + {1\over \bar V(r)}d\theta^{2}\eqn\dseven
$$
and the dilaton is
$$
D = -{1\over 2}\log(\eta(r)\sqrt{\bar V(r)}).\eqn\deight
$$
Thus it is possible to write down a non-linear $\sigma$-model starting
from an exactly solvable quantum mechanics problem. But there are a few
subtleties involved in showing this correspondence which are worth
pointing out. Firstly, one non-linear $\sigma$-model corresponds to a set
of quantum mechanical problems and secondly, the graviton and the dilaton
derived from the quantum mechanical problem should satisfy the
$\beta$-function equations for the $\sigma$-model to be conformally
invariant. This is an important constraint because the target space action,
which is an effective action derived from the $\beta$-function equations,
otherwise would not make sense.

\chapter{Summary and Discussion}

Once we have the full spectrum we can ask whether all the representations
are allowed. In unitary current algebra theories, only a finite number of
representations occur at a given level $k$. In unitary theories, $k$ is
always an integer, whereas, in our case, it can take any real value
because we are considering the coset model based on a non-compact group.
Allowed representations in unitary theories are called integrable
representations. The notion of integrable representations can also be
extended to the fractional levels of the current algebra\KW\MP. In the
case of a fractional level $k=t/u$ of the SU(2) current algebra, where $u$ is
a positive integer and $t$ is a non-zero integer with $u$ and $t$ coprime,
we get integrable representations if
$$
2u + t  - 2 \ge 0.\eqn\sd
$$
The black hole problem corresponds to the level $k = -9/4$ of the SU(2)
current algebra.(It is related to $k=9/4$ of the SL(2,R) by Wick rotation
of one of the coordinates.) It is easy to see that the black hole problem
does not satisfy the condition given in eq.\sd. Therefore all the
representations are non-integrable. Hence, we have no a priori reason to
rule out any representation.

We studied the black hole CFT by mapping the problem into a quantum
mechanical problem. We showed that this quantum mechanics problem can be
solved exactly. To show this, we used the techniques of SUSY quantum
mechanics and shape invariance. We determined both the bound state and
the scattering spectrum and identified it with the spectrum of the vertex
operators in the black hole CFT. These vertex operators are actually the
chiral algebra primary fields. We indicated how the remaining Virasoro
primary fields can be determined.

We also showed that the techniques of SUSY quantum mechanics and shape
invariance can be used for some CFTs with a $\sigma$-model
representation. Conversely, it is possible to show that for an exactly
solvable quantum mechanical problem we can associate a $\sigma$-model,
$i.e.$, it is possible to determine the background fields of the $\sigma$-model
starting from a specific solvable quantum mechanical problem. But this
correspondence between the quantum mechanical problem and the non-linear
$\sigma$-model involves a few subtle points. We hope to resolve them in
future.

{\bf Acknowledgements:} I am very grateful to Ashoke Sen for many illuminating
discussions and numerous suggestions. I would like to thank T. Jayaraman,
S. Rao, A. M. Sengupta and S. R. Wadia for useful discussions. I also thank
S. Rao and A. Sen for a critical reading of the manuscript.

\refout

\end